# AUTOMATED MORPHOLOGICAL ANALYSIS OF NEURONS IN FLUORESCENCE MICROSCOPY USING YOLOv8

Banan Alnemri and Arwa Basbrain

Department of Computer Science, Faculty of Computing and Information Technology King Abdulaziz University. Jeddah, Saudi Arabia

*ABSTRACT*

*Accurate segmentation and precise morphological analysis of neuronal cells in fluorescence microscopy images are crucial steps in neuroscience and biomedical imaging applications. However, this process is labor-intensive and time-consuming, requiring significant manual effort and expertise to ensure reliable outcomes. This work presents a pipeline for neuron instance segmentation and measurement based on a high-resolution dataset of stem-cell-derived neurons. The proposed method uses YOLOv8, trained on manually annotated microscopy images. The model achieved high segmentation accuracy, exceeding 97%. In addition, the pipeline utilized both ground truth and predicted masks to extract biologically significant features, including cell length, width, area, and grayscale intensity values. The overall accuracy of the extracted morphological measurements reached 75.32%, further supporting the effectiveness of the proposed approach. This integrated framework offers a valuable tool for automated analysis in cell imaging and neuroscience research, reducing the need for manual annotation and enabling scalable, precise quantification of neuron morphology.*

## 1. INTRODUCTION

Quantitative analysis of cellular structures is essential for biomedical research, diagnosis, and therapeutic development. Researchers can identify morphological changes related to a variety of diseases, such as cancer, neurodegenerative disorders, and infectious diseases, by precisely measuring cellular attributes such as size, morphology, and intracellular components. Regardless the substantial in biomedical image segmentation, neuron segmentation remains particularly challenging due to the overlapping cellular structures, and low-contrast boundaries in fluorescence microscopy images.Emphasizing the need for more robust and flexible methods for deep learning.One high-content, high-throughput imaging technique that is useful in explaining a variety of biological phenomena is the cell painting assay, which captures a wide range of cellular phenotypes in response to various perturbations[1].

Traditional approaches for cell measurement frequently use semiautomated image processing techniques along with manual annotation. However, according to[2] these methods are time-consuming, susceptible to interobserver variability, and have limited scalability. Recent advances in deep learning particularly in the domains of object detection and segmentation

models,provide a viable substitute for automated and high-throughput cellular analysis. Medical imaging applications have embraced the YOLO (You Only Look Once) model family due to its effectiveness and precision in real-time object detection[3].

In this work, a YOLOv8 was used to extract cellular measurements from a 2D fluorescent microscope image dataset that has already been created. The dataset used in this study is derived from an earlier biological work[4]. In that study, the neurogenic differentiation potential of stem cells isolated from immature wisdom teeth was investigated through culture and imaging. To enable a computational focus on neuron segmentation and morphological measurement, high-resolution fluorescence microscopy images from that study were repurposed to create a dedicated dataset for the current investigation. Through manual annotation, we produce ground truth measurements, which we then compare to the YOLOv8 predictions. The purpose of this assessment is to evaluate the accuracy and suitability of YOLOv8 for use in biomedical research, specifically in the analysis of cell morphology and the characterization of diseases. This study helps to improve the dependability of AI-driven image analysis methods by bridging the gap between deep learning models and practical medical applications.

## 2. RELATED WORK

The need for accurate, scalable, and repeatable quantification of cellular features has driven the growth of the field of automated cell analysis research. Cellular structures have historically been segmented and measured using conventional image processing methods like thresholding, watershed segmentation, and morphological filtering. However, these techniques frequently have difficulties with complicated biological images, which results in inconsistent segmentation outcomes.

Deep learning approaches have significantly improved the accuracy of cell detection and segmentation. In biomedical image segmentation tasks, CNN-based architectures like DeepLabV3+[5] and U-Net[6] have shown superior performance. These models used encoder-decoder structures to extract spatial features and generate precise segmentations. Additionally, object detection frameworks such as SSD[7] and Faster R-CNN[8] have been usedfor cell localization and counting tasks.

The work [3] first presented the YOLO architecture, which has since undergone several iterations that have increased detection accuracy and computational effectiveness.YOLOv8 provides a compelling trade-off between computational speed and accuracy and when compared to other segmentation architectures like U-Net and Mask R-CNN.Although U-Net is known for its accuracy in pixel-level segmentation, particularly in biomedical tasks, handling overlapping structures often calls for more post-processing and training time. Also Mask R-CNN is computationally more challenging and may have difficulties with fine neurite boundaries in high-density neuron images.YOLOv8 is ideally suited for high-resolution biological images since it integrates developments in object detection and instance segmentation.

YOLO models have been successfully used in earlier research to automatically detect cells in fluorescence and histopathological microscopy images, indicating their potential for a range of medical uses. However, there are still few thorough tests of YOLO-based models against high-quality ground truth data, which calls for more investigation into their applicability and dependability. Also, fewer studies have explored using YOLO models to extract precise morphological measurements such as cell length, width, area, and intensity, highlighting the novelty of the present work.

## 3. MATERIALS AND METHODS

This section outlines the dataset used in this study along with the deep learning-based methodology used for segmentation and morphological measurement.

### 3.1. Dataset

The dataset used in this study includes a total of 17,605 cells captured across 90 high-resolution 2D fluorescent microscope images (4084 × 4084 pixels each). The biological source of the data obtained from stem cells derived from the apical papilla of immature human wisdom teeth, which were then differentiated into neuron cells. Confocal fluorescence microscopy was used to clarify cellular morphology, that involve soma and neurite structures.

This dataset was designedto support the work of [4] in cell segmentation, quantification, and phenotype analysis. The dataset was previously created and annotated as part of earlier but not published yet work, and it is repurposed here to assess the measurement capabilities of the deep learning models. The dataset includes many challenges for segmentation as overlapping structures, variations in cell morphology. This repurposing emphasizes the potential for integrating deep learning with bioimage informatics for assisting high-throughput, automated morphological analysis in stem cell-based neuroscience research.

This paper's main contribution is assessing how well YOLOv8 can replicate these measurements and thereby support biomedical image analysis tasks.

### 3.2. Model

Before training the models, all input images were first converted to grayscale to simplify intensity-based analysis. Pixel values were normalized to the 0–255 range to ensure consistency across all images. The dataset was split into 70% for training, 15% for validation and 15% for testing.

YOLOv8 instance segmentation model was trained,Using the official Ultralytics framework implementation with default hyperparameters. For every object it detects, the model produces instance masks and bounding boxes. For each detected instance a binary mask from the YOLOv8 predictions was extracted. In a similar manner, a custom Python script that made use of the numpy and opencv-python libraries was used to convert manually annotated ground truth masks from V7 Darwin from JSON format to binary masks.

The morphological measurements for each segmented objectincluding length, width, area, and intensitywere computed from both ground truth and prediction as follows:

#### 3.2.1. Cell Length and Width

The height of each object's bounding box was used to estimate its length and width. We obtained (x, y, width, height) for each mask using an OpenCV's function, then we noted the height as the cell length and the width as the cell width. This method guaranteed consistent length and width estimation across irregularly shaped cells.

### 3.2.2. Cell Area

Using NumPy's function, the area was calculated by counting the number of non-zero (foreground) pixels in each binary mask. The pixel count reflects the number of pixels belonging to the cell instance within the mask.

### 3.2.3. Cell Intensity

The binary mask was used as an index on the original grayscale image to extract intensity values. We gathered all pixel intensity values inside the mask region using NumPy. Three NumPy's functions were used to calculate the mean, minimum, and maximum from this pixel array.

To ensure consistency between training and evaluation images, all measurements were calculated using the 640×640 resized resolution. Intensity values were reported on the 0–255 grayscale scale, length and width in pixels, and area in square pixels. Measurements from all detected instances (both in ground truth and prediction) were combined for each image. The ground truth measurements were therefore calculated directly from the manual annotations that were exported from V7 Darwin. They were then converted into binary masks and examined using the same functions that were used to analyze the predictions. This eliminated methodological bias by guaranteeing that the measurement pipelines for YOLOv8 predictions and ground truth were identical.

To visually clarify the measurement methodology, Figure 1. depicts the extraction of morphological parameters from each segmented cell. The figure illustrates a simplified representation of a segmented neuron-like cell. The arrows denote the measurement of length and width along the longest and shortest axes, respectively. The area applies to the aggregate pixel count within the binary mask, whereas intensity is obtained from the grayscale pixel values present in that region.

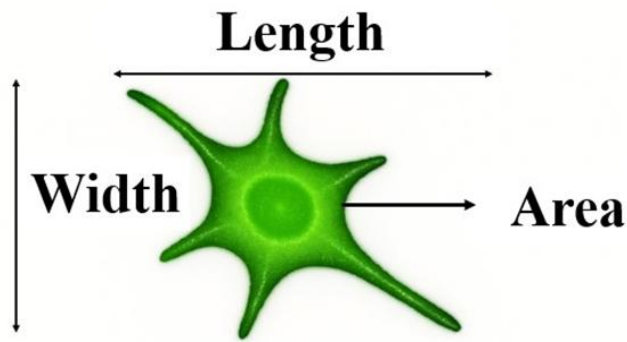

Figure 1. Illustration of the measurement of morphological characteristics on a segmented neuron-like cell.

## 4. RESULTS

Throughout the test dataset, two main evaluation criteria were used to evaluate the YOLOv8 model's performance: segmentation performance and measurement accuracy. When combined, these enable us to assess the model's ability to accurately reproduce biologically significant measurements like area, length, and intensity in addition to its ability to recognize neuron instances.

To evaluate segmentation quality, we used the following standard metrics: Precision to measure the proportion of correctly predicted cells out of all predictions made. Recall measuring the proportion of correctly predicted cells out of all actual cells. The F1 Score represents a balance between precision and recall. IoU (Intersection over Union) to evaluate how well the predicted masks overlap with the ground truth masks and Panoptic Quality (PQ): PQ = SQ * RQ. Where SQ assesses mask-level similarity (IoU), and RQ evaluates the detection correctness of instances. The model showed an overall accuracy of 97%. Table 1. shows the segmentation metrics results.

Table 1. Results of YOLOv8 model segmentation evaluation.

| | |
|---|---|
| **Precision** | 98.17% |
| **Recall** | 99.50% |
| **F1 Score** | 98.82% |
| **IoU Accuracy** | 97.69% |
| **PQ** | 97.69% |
| **SQ** | 97.69% |
| **RQ** | 100.00% |
| **Overall Accuracy** | 97.72% |

To evaluate the accuracy of the predicted morphological measurements (length, width, area, and intensity values), we compared each predicted value with its corresponding ground truth. For every metric, the procedure was repeated for every matched cell in the test set. The average accuracy for each metric, including length, width, area, and grayscale intensity values (min, mean, and max), was then calculated by averaging these accuracy scores.

In addition to numerical evaluation, Figure 2. provides a visual comparison of the original fluorescence microscopy image, manually annotated ground truth mask, and YOLOv8 prediction. Displaying the original input, ground truth segmentation with labeled cells, and predicted segmentation by YOLOv8. By demonstrating the precision of cell boundary detection and the propensity for over-detection in grouped regions, this visualization supports the quantitative results.

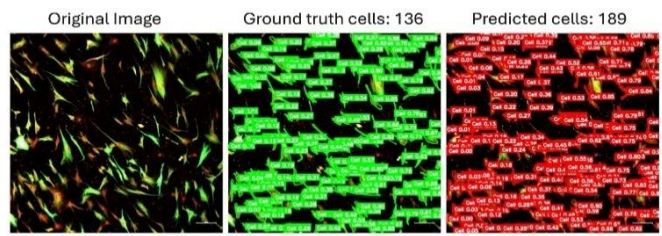

Figure 2. Visual comparison between original image, ground truth segmentations, and YOLOv8 predictions.

## 5. DISCUSSION

To justify the choice of YOLOv8 as the primary segmentation model for this study, we performed a baseline comparison against two instance segmentation architectures: Mask R-CNN and YOLOv11. YOLOv8 outperformed the other models across the evaluation metrics. This work then demonstrated the efficiency of a YOLOv8-based segmentation pipeline in extracting biologically significant measurements from 2D fluorescence microscopy images of stem-cell-derived neurons. The model achieved high segmentation accuracy, confirming its robustness in detecting and segmenting individual neuronal structures, even in dense and morphologically diverse fields.

To improve the understanding of the YOLOv8 model's efficacy in delineating cellular morphology, we computed the measurement-specific accuracy for each extracted feature: length, width, area, and intensity (minimum, mean, and maximum). As shown in Table 2, the model confirmed a strong ability to preserve brightness characteristics by achieving high accuracy in estimating both the mean intensity (88.40%) and maximum intensity (99.62%). Morphological dimensions with accuracies of 82.98% and 82.08%, respectively, were also accurately predicted. Due to the fluctuating low-signal regions in fluorescence microscopy, minimum intensity estimation had the lowest accuracy (22.15%) and area estimation performed marginally worse (78.66%). Overall, the model demonstrated reliable performance in quantifying most morphological features while highlighting important areas for additional development, with an average accuracy of 75.32% across all measurements.

Table 2. Accuracy of YOLOv8 predictionscompared to ground truth.

| Measurement | Accuracy |
| --- | --- |
| Length | 82.98% |
| Width | 82.08% |
| Area | 78.66% |
| Min Intensity | 22.15% |
| Mean Intensity | 88.40% |
| Max Intensity | 99.62% |
| Overall accuracy | 75.32% |

In addition to performing dependable segmentation, the suggested pipeline makes it easier to extract morphometric features, which are frequently employed in cell biology, neuroscience, and medical image analysis. There are still limitations like how well it works with long or overlapping cells. Despite that, the model occasionally produced over-segmented results especially in dense regions with overlapping cells, which leads to inflated cell counts and reduced precision. This can be caused by the complex cell's morphology and variations in fluorescence intensity. On the other hand, some small cells were missed entirely, especially in low-signal areas causing false negatives and lower recall. Although there are no post-processing steps in this study but future work might investigate transformer-based instance segmentation models to increase accuracy.

## 6. CONCLUSIONS

This work provided a comprehensive framework for neuron segmentation and morphological analysis using YOLOv8 applied to 2D fluorescence microscopy images. The model showed outstanding segmentation performance and produced precise measurements of cell morphology, such as length, area, and grayscale intensity. The predicted measurements demonstrated a strong correlation with manually annotated ground truth data, despite small differences. The suggested pipeline is appropriate for real-time applications in biomedical imaging since it successfully integrates segmentation and quantitative analysis into a single model. Expanding the dataset, improving segmentation for intricate structures, and incorporating biological validation metrics are possible future enhancements.

## ACKNOWLEDGEMENTS

The author would like to thank the Department of Computer Science at King Abdulaziz University for providing the research environment and support.

## AUTHORS

**Banan Alnemri** is a master's student in the Department of Computer Science at King Abdulaziz University, Jeddah, Saudi Arabia. Her research interests include computer vision, biomedical image analysis, and deep learning. She is currently focusing on developing AI-based solutions for medical imaging tasks, particularly in neuron segmentation and morphological analysis using fluorescence microscopy data.

**Dr. Arwa M. Basbrain** received her Ph.D. in Computer Science specializing in facial recognition from the University of Essex in 2021. She is currently an Assistant Professor at King Abdulaziz University.Her research interests include artificial intelligence, computer vision, deep learning, health informatics, and natural language processing. She is also leading a project titled Intelligent Search Engine for the Holy Quran, which focuses on natural language processing applications. Her teaching focuses on AI-related subjects such as neural networks and computer vision.